\documentclass[reprint,aps,prb,english,superscriptaddress]{revtex4-2}

\usepackage{graphicx}
\usepackage{booktabs}
\usepackage{bm,amssymb,amsmath,mathrsfs,amstext,latexsym,physics}
\usepackage[T1]{fontenc}

\usepackage[usenames,dvipsnames]{xcolor}
\usepackage[colorlinks=true,citecolor=blue,urlcolor=blue,linkcolor=blue]{hyperref}
\usepackage[normalem]{ulem}

\usepackage{soul}
\setstcolor{red}

\begin{document}

\title{Quantifying when hyperuniformity of a many-particle system leads to \\
uniformity across length scales}

\author{Carlo Vanoni}
\affiliation{Department of Physics, Princeton University, Princeton, New Jersey, 08544, USA}

\author{Paul J. Steinhardt}
\affiliation{Department of Physics, Princeton University, Princeton, New Jersey, 08544, USA}

\author{Salvatore Torquato}
\affiliation{Department of Chemistry, Princeton University, Princeton, New Jersey 08544, USA}
\affiliation{Department of Physics, Princeton University, Princeton, New Jersey, 08544, USA}
\affiliation{Princeton Institute for the Science and Technology of Materials, Princeton University, Princeton, New Jersey 08544, USA}
\affiliation{Program in Applied and Computational Mathematics, Princeton University, Princeton, New Jersey 08544, USA}

\date{\today}

\begin{abstract}
    Hyperuniform systems are distinguished by an unusually strong suppression of large-scale density fluctuations and, consequently, display a high degree of uniformity at the largest length scales. 
    In some cases, however, enhanced uniformity is expected to be present even at intermediate and, possibly, down to small length-scales.
    There exist three different classes of hyperuniform systems, where class I and class III are the strongest and weakest forms of hyperuniformity, respectively.
    We utilize the local number variance $\sigma_N^2(R)$ associated with a window of radius $R$ as a diagnostic to quantify the approach to the asymptotic large-$R$ hyperuniform scaling of a variety of class I, II, and III hyperuniform systems. 
    We find, for all the class I systems we analyzed, which include crystals, quasicrystals, disordered stealthy hyperuniform systems, and the one-component plasma, a faster approach to the asymptotic scaling of $\sigma_N^2(R)$, governed by corrections with integer powers of $1/R$.
    Thus, we conclude that this represents the highest degree of effective uniformity from small to large length scales among hyperuniform systems.
    Class II hyperuniform systems, such as Fermi-sphere point processes, are characterized by logarithmic $1/\ln(R)$ corrections to the asymptotic scaling and consequently, a lower degree of local uniformity compared to class I. 
    Class III hyperuniform systems, such as perturbed lattice patterns, present an asymptotic scaling of $1/R^{\alpha}$, $0 < \alpha < 1$, implying, curiously, an intermediate degree of local uniformity compared to class I and II.
    In addition, our study provides insight into when experimental and numerical finite systems are representative of large-scale behavior.
    Our findings may thereby facilitate the design of hyperuniform systems with enhanced physical properties, such as transport and mechanical properties, arising from local uniformity.
\end{abstract}

\maketitle

\section{Introduction}

Hyperuniform many-particle systems in $d$-dimensional Euclidean space $\mathbb{R}^d$ are characterized by an anomalous suppression of large-scale density fluctuations and provide a unified framework to classify crystals, quasicrystals, and special disordered systems~\cite{To03a,To18a}. 
Examples of disordered hyperuniform systems include perfect glasses~\cite{Zh16a}, sphere packings and maximally random jammed states~\cite{Dr15,Za11c,To00b,Ma23}, jammed athermal soft-sphere models of granular media~\cite{Si09,Be11}, jammed thermal colloidal packings~\cite{Wi20,Ma20a,Ni21,Ku11}, one-component plasmas~\cite{Le00} and ground state wave function associated with the fractional quantum Hall eﬀect~\cite{Laugh87}, critical absorbing states of random organization models~\cite{He15,Ma19,Wiese2024Hyperuniformity,Hazra_2025}, random matrices and number theory~\cite{To08b,Sc09,Mon73}, classical and quantum spin liquids~\cite{Chen2025Anomalous}.
There are also many experimentally relevant applications of hyperuniformity, such as avian photoreceptor patterns~\cite{Ji14}, receptor organization in the immune system~\cite{Ma15}, jammed bidisperse emulsions~\cite{Ri17}, certain medium- or high-entropy alloys~\cite{Chen2023Multihyperuniformity}, amorphous carbon nanotubes~\cite{Chen2022Disordered}, certain metallic glasses~\cite{Zhang2023Approach}. 
Furthermore, disordered hyperuniform materials are receiving great attention due to their novel physical properties with advantages over their crystalline counterparts~\cite{florescu_designer_2009,Zi15b,Zh16,Ma16,De16,leseur_highdensity_2016,froufe-perez_band_2017,gkantzounis_hyperuniform_2017,Zh19,Ro19,zhou_ultrabroadband_2020,Ae22,To22b,alhaitz_experimental_2023,merkel_stealthy_2023,kim_effective_2023,Kim_2024_extraordinary,kim_accurate_2024,Vanoni2025Dynamical}.

A hyperuniform system in $\mathbb{R}^d$ is defined by a number variance $\sigma^2_N(R)$ of particles within a spherical observation window of radius $R$, $\sigma^2_N(R) \equiv \langle N^2 \rangle - \langle N \rangle^2$, that {\it globally} grows~\footnote{By global growth, we mean up to small-scale fluctuations, {\it i.e.,} of the order of the nearest neighbor-distance. In fact, in general, the coefficient $B_N(R)$ in Eq.~\eqref{eq:sigma_series} can depend on $R$, and in the case of periodic point configurations, it is bounded and fluctuates around a constant~\cite{To03a,To18a}.} 
for large $R$ more slowly than the window volume $R^d$ ({\it i.e.}, slower than the growth law for typical disordered systems).
Three different hyperuniform variance scalings or {\it classes} that span between window surface-area $R^{d-1}$ and volume growth $R^d$ are possible \cite{To18a}; see Sec. \ref{background} for details.
Equivalently, hyperuniformity can be defined in momentum space for all crystals and disordered hyperuniform systems as the vanishing of the structure factor $S(\mathbf{k})$ as the wave number  $k=|\mathbf{k}|$ goes to zero, {\it i.e.}, $\lim_{|\mathbf{k}| \to 0} S(\mathbf{k}) = 0$.

Thus, hyperuniformity quantifies the degree of enhanced {\it uniformity} at the largest length scales of a system,
as measured by the anomalous suppression of density fluctuations, relative to that of ordinary disordered systems, such as typical liquids and glasses.
As a measure of density fluctuations within finite observation windows, the local number variance offers a natural framework for assessing uniformity from microscopic to macroscopic scales.
A fascinating question is whether such uniformity in hyperuniform systems is heralded at smaller length scales? While we know that some previous results suggest that certain crystals~\cite{To03a,Za09,To18a} and quasicrystals~\cite{Li17a,Og17} signal hyperuniformity at short and intermediate length scales via the behavior of the local number variance, this question has not been systematically studied via this metric.
Recently, Maher and Torquato~\cite{Maher2024Local} used the local number variance as a local {\it order metric} in a variety of different models of antihyperuniform, hyperuniform, and ordered hyperuniform many-particle systems, showing that it can sensitively describe the degree of order/disorder across length scales.

In the present work, we repurpose the scaled local number variance used in Ref.~\cite{Maher2024Local} to determine when and to what extent the enhanced uniformity at the largest length scale of two-dimensional (2D) and three-dimensional (3D) hyperuniform models of class I, II, and III is inherited at smaller scales.
Specifically, we aim to address the approach of the number density fluctuations to their asymptotic behaviors, driven by the broad relevance of this topic across diverse systems and applications. 
For this purpose, we exactly analyze the scaled local number variance, defined for class I and III systems as $\sigma_N^2(R)/R^{\gamma_d}$ for all $R$, being $\sigma_N^2(R) \sim R^{\gamma_d}$ for $R \to \infty$, while, for class II systems, a logarithmic correction needs to be added, $\sigma_N^2(R)/(R^{\gamma_d}\ln R)$ for all $R$, being $\sigma_N^2(R) \sim R^{\gamma_d} \ln R$ for $R \to \infty$. These relations define the $d$-dependent exponent $\gamma_d$ that governs the infinite distance behavior.
Importantly, we will show in Sec.~\ref{sec:classII} that the different scaling behavior that class II systems present will lead to the weakest form of local uniformity among hyperuniform systems. 
By studying values of $R$ for which the cumulative running average of the scaled local number variance attains its infinite-distance value, we will determine which microstructures are effectively uniform from small to large length scales.

The answer to this fundamental question also has practical implications. 
For example, since all numerically and experimentally generated many-particle systems are necessarily finite, it is of great interest to be able to ascertain at which length scales it suffices to deem that a large but finite system is hyperuniform, since the latter requires an infinitely large system, in principle. We also expect that enhanced local uniformity can lead to improved physical properties. For instance, in the context of mechanical properties, it has been shown~\cite{To00a} that the suppression of volume-fraction fluctuations in disordered hyperuniform particulate composites is known to suppress crack propagation within the matrix phase.  Advantages are expected to be present also in transport and optical properties, as suggested by the improved characteristics of stealthy hyperuniform systems~\cite{kim_effective_2023,kim_accurate_2024,Vanoni2025Dynamical,To21d}. For example, the local uniformity can lead to beneficial effects in the diffusion spreadability at intermediate time scales.

We note that, in a recent series of works~\cite{Salvalaglio2024Persistent,Milor2025Inferring}, Salvalaglio et al. have employed tools coming from topology, in particular persistent homology, to address the influence of global hyperuniformity on local structure. Persistent homology is a scale-free tool and therefore well-suited for such analyses. The emphasis in Refs.~\cite{Salvalaglio2024Persistent,Milor2025Inferring}, however, was on topological properties most pronounced at the nearest-neighbor scale, whereas our approach is specifically designed to probe local uniformity systematically across all length scales.

The remainder of the manuscript is organized as follows. In Sec.~\ref{background} we introduce some background notions that we will use to obtain our results. 
In Sec.~\ref{sec:classI} we discuss class I hyperuniform systems, and we consider lattices in Sec.~\ref{sec:lattices}, quasicrystals in Sec.~\ref{sec:quasicrystal}, patterns generated from the $g_2$-invariant step function in Sec.~\ref{sec:g2invariant}, disordered stealthy hyperuniform ground states in Sec.~\ref{sec:SHU}, one-component plasma in Sec.~\ref{sec:onecompplasma} and Gaussian $g_2$ in Sec.~\ref{sec:gaussian}.
We then address class II hyperuniform systems in Sec.~\ref{sec:classII}, focusing on Fermi sphere point processes in Sec.~\ref{sec:Fermi}. We then discuss class III hyperuniform systems in Sec.~\ref{sec:classIII}, examining perturbed lattice point patterns in Sec.~\ref{sec:perturbed}. We then discuss our results in Sec.~\ref{sec:conclusions}.
In Fig.~\ref{fig:patterns} we report a representative two-dimensional configuration for each of the models that we consider in this paper.
We devote the supplemental material to the study of the scaled number variance in non-hyperuniform and anti-hyperuniform systems.

\section{Background and Methods}
\label{background}

\subsection{Number variance}

\begin{figure*}
    \centering
    \includegraphics[width=0.3\linewidth]{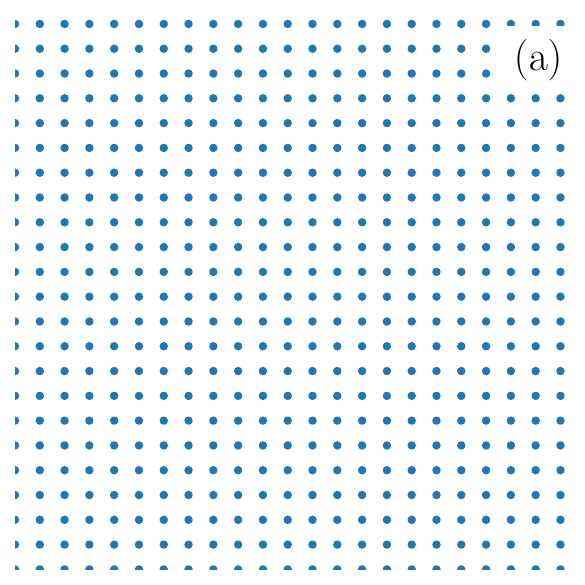}
    \includegraphics[width=0.3\linewidth]{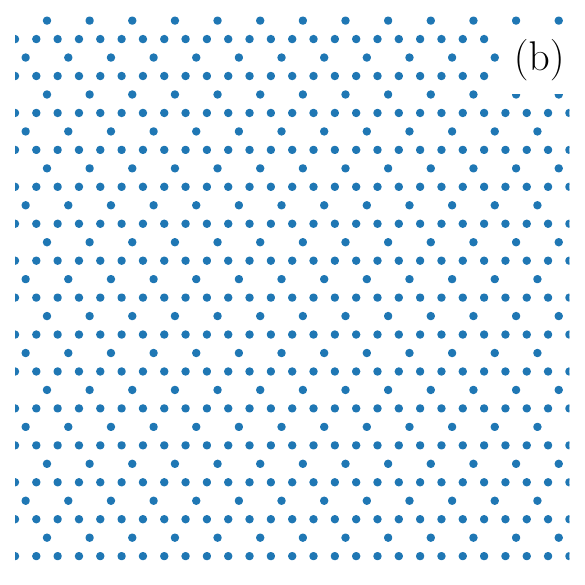}
    \includegraphics[width=0.3\linewidth]{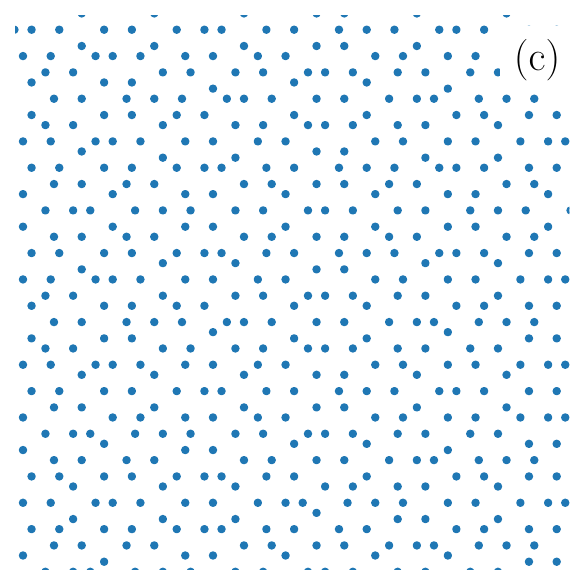}
    \includegraphics[width=0.3\linewidth]{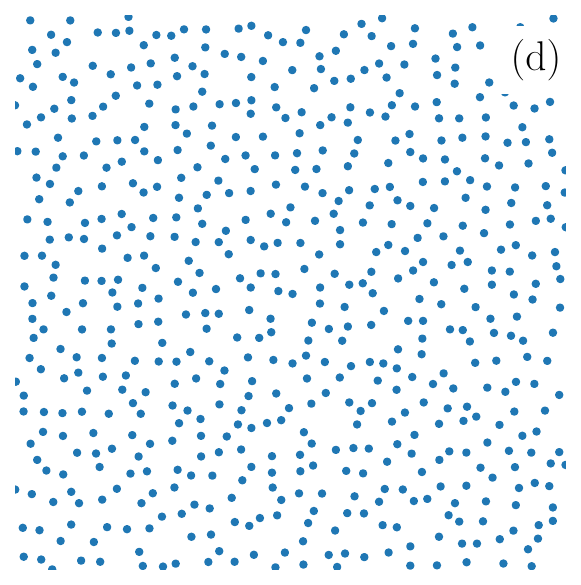}
    \includegraphics[width=0.3\linewidth]{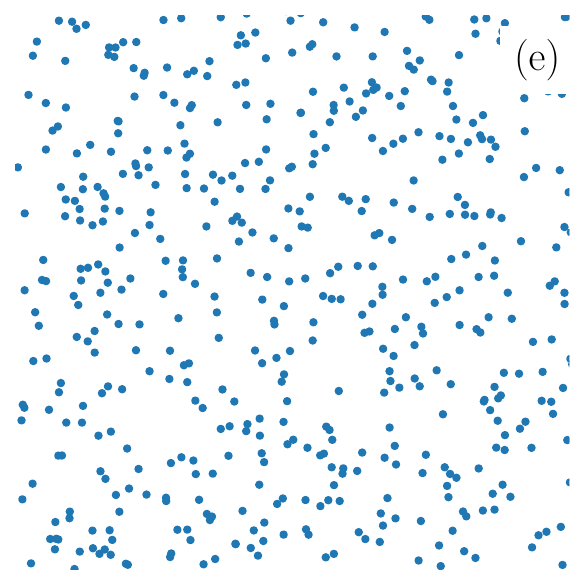}
    \includegraphics[width=0.3\linewidth]{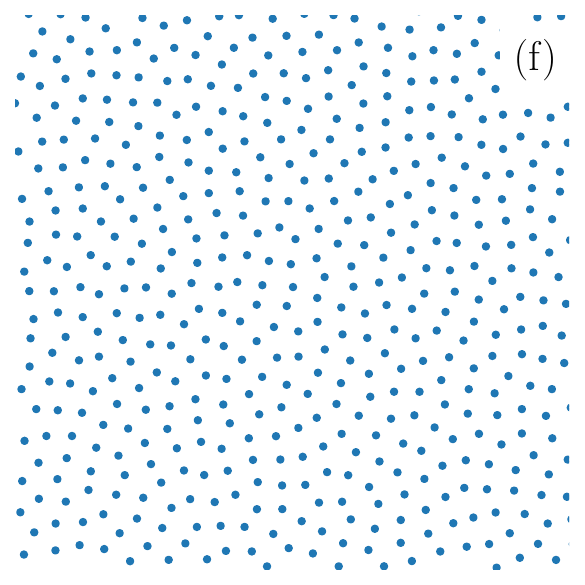}
    \includegraphics[width=0.3\linewidth]{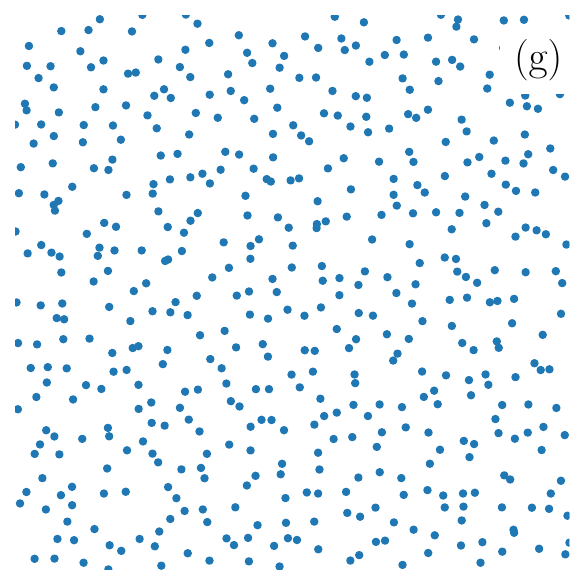}
    \includegraphics[width=0.3\linewidth]{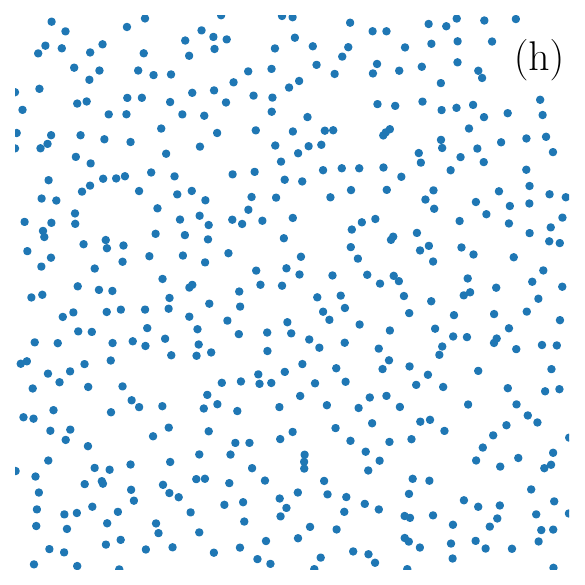}
    \includegraphics[width=0.3\linewidth]{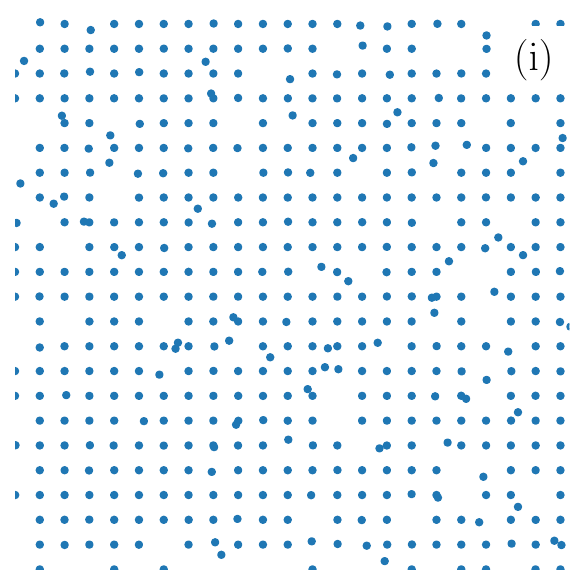}
    \caption{Two-dimensional representative configurations for each of the models that we consider at unit number density: 
    \\(a) square lattice, (b) kagome lattice, (c) Penrose tiling quasicrystal, (d) $g_2$-invariant step function, (e) stealthy hyperuniform ground state with $\chi = 0.0025$, (f) stealthy hyperuniform ground state with $\chi = 0.45$, (g) one-component plasma, (h) Fermi-sphere point process, and (i) perturbed lattice with $\alpha = 0.25$ and $\delta = 0.0001$.}
    \label{fig:patterns}
\end{figure*}

We briefly review some basic concepts about the local number variance that are germane to this study.
Consider a point pattern in $d$-dimensional Euclidean space $\mathbb{R}^d$. At position $\mathbf{x}_0 \in \mathbb{R}^d$, consider a spherical window of radius $R$ and let $N(\mathbf{x}_0;R)$ be the number of points contained in the window. Denoting with $\langle N(R) \rangle = \rho \, v_1(R)$ the ensemble average of $N(\mathbf{x}_0;R)$, the number variance is simply given by
\begin{equation}
    \sigma_N^2(R) \equiv \langle N(R)^2 \rangle - \langle N(R) \rangle^2.
\end{equation}

It is possible to express the number variance in terms of the pair correlation function $g_2(\mathbf{r})$, or equivalently, the total correlation function $h({\bf r}) = g_2({\bf r})-1$~\cite{To03a,To18a}. 
In particular, the local number variance $\sigma^2_N(R)$ at number density $\rho$ for a spherical window of radius $R$ is given, in direct space representation, by
\begin{equation}
\label{local-1}
    \sigma^2_N(R) = \rho v_1(R) \left[ 1+ \rho \int_{\mathbb{R}^d} h({\bf r}) \alpha_2(r;R) d {\bf r} \right],
\end{equation}
where $r = |{\bf r}|$, $v_1(R) = \pi^{d/2} R^d/ \Gamma(1+d/2)$ is the volume of a spherical window of radius $R$ in $d$ dimensions, and $\alpha_2(r;R) = v_2^{\mathrm{int}}(r; R)/v_1(R)$ is the scaled intersection volume of two spheres of radius $R$ with centers at distance $r$, whose expression is given by~\cite{To03a}
\begin{equation}
\label{eq:alpha2}
    \alpha_2(r;R) = \frac{2 \Gamma(1+d/2)}{\sqrt{\pi} \Gamma[(d+1)/2]} \int_0^{\cos^{-1}(r/2R)} \sin^d(\theta) d\theta.
\end{equation}
One can explicitly expand Eq.~\eqref{eq:alpha2} in odd powers of $r/2R$~\cite{To06b}: for example, to first order one gets~\cite{To03a}
\begin{equation}
    \alpha_2(r;R) = 1-c(d) \frac{r}{2R} + o\left( \frac{r}{2R} \right),
\end{equation} 
with the coefficient $c(d)$ given by
\begin{equation}
\label{eq:cofd}
    c(d) = \frac{2\Gamma(1+d/2)}{\pi^{1/2} \Gamma[(1+d)/2]}.
\end{equation}
Using this expansion, one obtains~\cite{To03a}
\begin{align}
\label{eq:sigma_series}
     \sigma_N^2(R) = 2^d \phi & \left[ A_N(R) \left( \frac{R}{D} \right)^d + B_N(R) \left( \frac{R}{D} \right)^{d-1} \right. \nonumber\\
    &\left. + o\left( \left( \frac{R}{D} \right)^{d-1} \right) \right] \quad (R \to \infty),
\end{align}
where $D$ is a microscopic length scale ({\it e.g.}, the mean nearest-neighbor distance), and $\phi = \rho v_1(D/2)$ is a dimensionless density.
The coefficient $A_N(R)$ is a $d$-dependent asymptotic coefficient that multiplies terms proportional to the window volume and is obtained from the total correlation function as
\begin{equation}
    A_N(R) = 1 + \frac{\phi}{v_1(D/2)} \int_{|\mathbf{r}| \leq 2R} h(\mathbf{r}) \, d\mathbf{r}.
\end{equation}
The coefficient $B_N(R)$ represents the window surface area and is given by
\begin{equation}
    B_N(R) = -\frac{\phi \, c(d)}{2D \, v_1(D/2)} \int_{|\mathbf{r}| \leq 2R} h(\mathbf{r}) |\mathbf{r}| \, d\mathbf{r},
\end{equation}
with the coefficient $c(d)$ is given by Eq.~\eqref{eq:cofd}.
In the $R \to \infty$ limit, it is easy to show that the volume coefficient $A_N(R)$ is equal to the structure factor $S(\mathbf{k})$ at zero  wave number, {\it i.e.},
\begin{equation}
\label{eq:A_N}
    \overline{A}_N \equiv \lim_{R \to \infty} A_N(R) = \lim_{|\mathbf{k}| \to 0} S(\mathbf{k}) \geq 0.
\end{equation}

The structure factor $S(\mathbf{k})$ is related to the Fourier transform of the total correlation function $\tilde{h}({\bf k})$ via $S(\mathbf{k}) = 1 + \rho \tilde{h}({\bf k})$. As a consequence, Eq.~\eqref{local-1} can be equivalently expressed in Fourier space, and it reads
\begin{equation}
\label{eq:sigma_F_space}
    \sigma_N^2(R) = \rho v_1(R)\left[\frac{1}{(2\pi)^d} \int_{\mathbb{R}^d} S(k) \tilde{\alpha}_2(k;R) d {\bf k}\right]
\end{equation}
where 
\begin{equation}
    \tilde{\alpha}_2(k;R) = 2^d \pi^{d/2} \Gamma(1+d/2)\frac{[J_{d/2}(kR)]^2}{k^d}.
\end{equation}
In statistically isotropic point patterns, the pair statistics are functions only of the radial distance $r$. It is then useful to recall that, for radial functions $f(\mathbf{r}) \equiv f(r)$, the integral over the volume can be written as an integral over the radial direction as
\begin{equation}
    \int_{\mathbb{R}^d} f(\mathbf{r})\, d {\bf r} = \frac{2 \pi^{d/2}}{\Gamma(d/2)}\int_0^{\infty}f(r)\, r^{d-1} dr.
\end{equation}

\subsection{Hyperuniformity}

A hyperuniform system is such that the large-scale density fluctuations are suppressed compared to those of garden-variety disordered systems. In the latter, for large $R$, $\sigma_N^2(R) \sim R^d$, whereas in hyperuniform systems, by definition,
\begin{equation}
\label{eq:def_hyp}
    \lim_{R\to \infty} \frac{\sigma_N^2(R)}{R^d} = 0.
\end{equation}
For perfect crystals and disordered hyperuniform systems, the definition Eq.~\eqref{eq:def_hyp} is equivalent, in Fourier space, to the vanishing of the structure factor at zero wave number, {\it i.e.}, $\lim_{|\mathbf{k} \to 0|}S(\mathbf{k}) = 0$. In particular, if the structure factor vanishes as a power-law \begin{equation}
\label{eq:Skalpha}
    S(\mathbf{k})\sim |\mathbf{k}|^{\alpha}, 
\end{equation}
the asymptotic behavior at large $R$ of the number variance is controlled by the exponent $\alpha$, and we can distinguish three different cases
\begin{align}
\label{eq:classes}
\sigma^2_N(R) \sim
\begin{cases}
    R^{d-1},  & \text{if } \alpha > 1 \quad\text{(Class I)},\\
    R^{d-1} \ln R,  & \text{if } \alpha = 1 \quad\text{(Class II)},\\
    R^{d-\alpha}, & \text{if } 0 < \alpha < 1 \quad \text{(Class III)}.
\end{cases}
\end{align}
It is then clear that $\gamma_d = d-1$ in classes I and II, while $\gamma_d = d-\alpha$ in class III.
From Eq.~\eqref{eq:A_N}, it is clear that hyperuniform systems are such that $\overline{A}_N = 0$. Moreover, in class I systems, the coefficient $B_N(R)$ saturates to a constant value in the large-$R$ limit, and therefore
\begin{equation}
    \sigma_N^2(R) \sim 2^d \,\phi \, \overline{B}_N \left(\frac{R}{D}\right)^{d-1}, \quad (R \to \infty),
\end{equation}
being $\overline{B}_N = \lim_{R \to \infty} B_N(R)$.

A special set of hyperuniform systems goes under the name of stealthy hyperuniformity. 
In such systems, the structure factor vanishes for a range of wave numbers around the origin, $S(\mathbf{k}) = 0$ for $0 < k \leq K$, and consequently, perfect crystals are stealthy systems.
Clearly, stealthy hyperuniform systems belong to class I.

In the following, we will consider the radial distance dependence of the number variance. For large $R$, the asymptotic scaling in governed by a $d$-dependent exponent $\gamma_d$ such that $\sigma_N^2(R) \sim R^{\gamma_d}$. 
In  the case of  crystals and quasicrystals (see Sec.~\ref{sec:lattices} and \ref{sec:quasicrystal}), however, the limit $\lim_{R \to \infty} \sigma_N^2(R) / R^{\gamma_d}$ might not exist because $\sigma_N^2(R) / R^{\gamma_d}$ possesses small-scale oscillations~\cite{To03a,Za09,Li17a}. 
This is a consequence of their order and symmetries, but does not imply that they lack uniformity across length scales. This observation motivates the repurposing of the cumulative moving average of the scaled number variance \cite{To03a}
 \begin{equation}
 \label{eq:Lambda}
    \overline{\Lambda}(L) \equiv \frac{1}{L}\int_{0}^L \frac{\sigma_N^2(R)}{(R/D)^{\gamma_d}} dR,
\end{equation}
and we denote with $\overline{\Lambda}$ the limit $\overline{\Lambda} \equiv \lim_{L\to \infty} \overline{\Lambda}(L)$.
The cumulative moving average thus isolates the rate of convergence of the scaled number variance, especially in the presence of small-scale oscillations.

Let us introduce the scaled number variance $\Sigma^2(R)$, which we will use in this paper to characterize uniformity across length scales in hyperuniform point patterns
\begin{equation}
\label{eq:Sigma_def}
    \Sigma^2(R) \equiv 
    \begin{cases}
        \frac{\sigma_N^2(R)}{(R/D)^{\gamma_d} \overline{\Lambda}} \quad \text{for classes I and III,} \\
        \frac{\sigma_N^2(R)}{(R/D)^{\gamma_d} \ln(R/D) \overline{\Lambda}}\quad \text{for class II}.
    \end{cases}
\end{equation}
By construction, in the large-$R$ limit, $\Sigma^2(R) = 1$ or oscillates around unity, and can therefore be written as
\begin{equation}
    \Sigma^2(R) = 1 - F(R),
\end{equation}
with $F(R)$ averaging to zero at large distances. The dependence on $R$ of $\Sigma^2(R)$, and thus $F(R)$, will provide information about the uniformity of the model on short and intermediate scales, in the approach to the large-scale hyperuniformity.

\section{Results -- Class I}
\label{sec:classI}

In this Section, we investigate the local number variance of Hyperuniform systems belonging to Class I, namely having $\sigma_N^2(R) \sim R^{d-1}$. 
In these systems, the scaled number variance $\Sigma^2(R)$, defined in Eq.~\eqref{eq:Sigma_def}, takes the form~\cite{To03a}
\begin{equation}
    \Sigma^2(R) = 1+ \frac{A_N(R)(R/D) + (B_N(R) - \overline{B}_N)}{\overline{B}_N} + o(1).
\end{equation}
We will determine the scaled number variance Eq.~\eqref{eq:Sigma_def} for lattices (Sec.~\ref{sec:lattices}), quasicrystals (Sec.~\ref{sec:quasicrystal}), $g_2$-invariant unit step function processes (Sec.~\ref{sec:g2invariant}), disordered stealthy hyperuniform systems (Sec.~\ref{sec:SHU}), one-component plasma (Sec.~\ref{sec:onecompplasma}) and systems with Gaussian correlations (Sec.~\ref{sec:gaussian}). In most cases, we will provide expressions valid in any dimensions, specializing in 2D and 3D.

\subsection{Lattices}
\label{sec:lattices}

\begin{figure}
    \centerline{\includegraphics[width=3in,keepaspectratio,clip=]{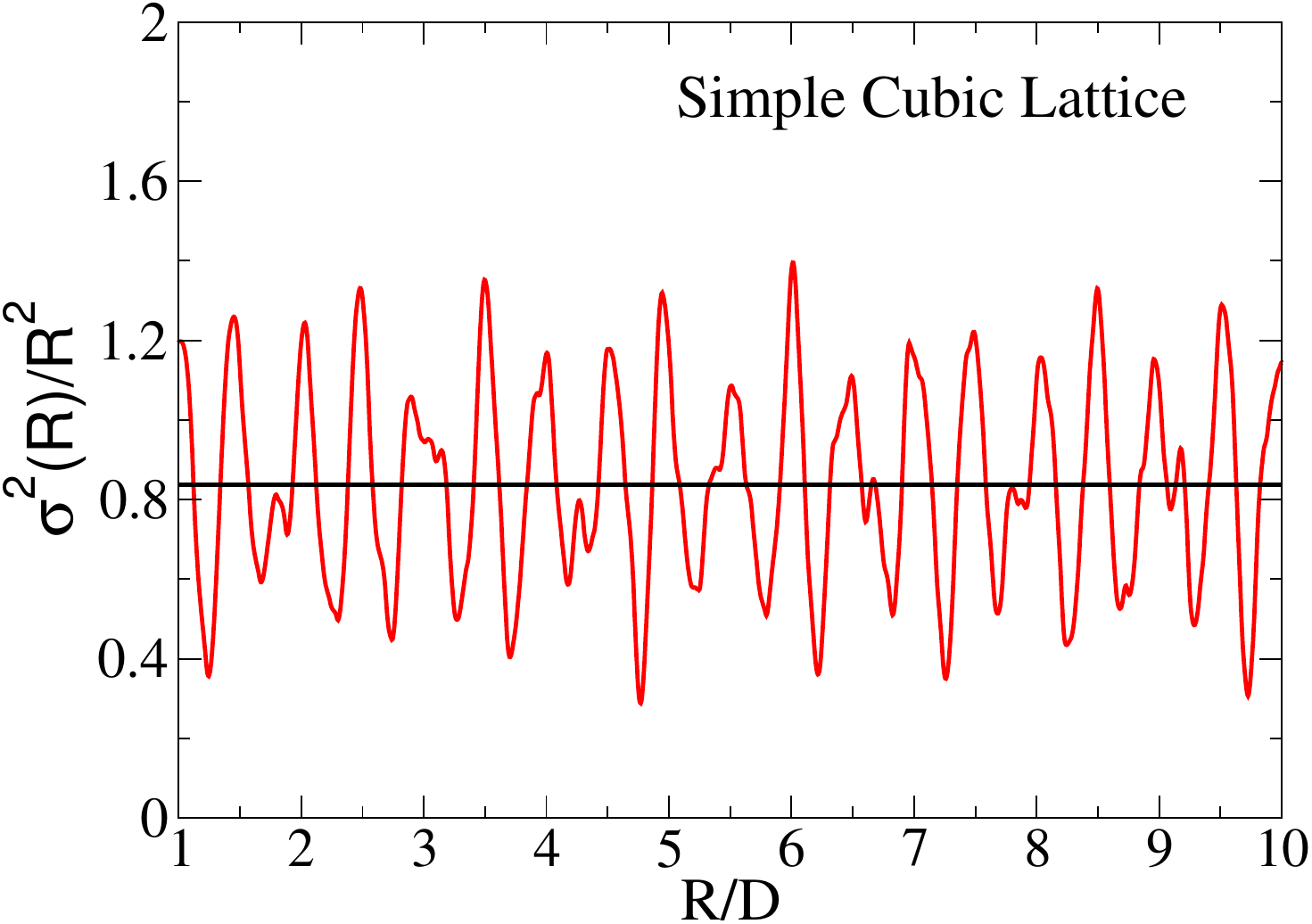}}
    \caption{ The scaled local variance $\sigma_N^2(R)/R^2$ versus $R/D$ for the simple cubic lattice, where $D$ is the nearest-neighbor distance,  and we take $\phi=\pi/6\approx 0.5236$. The average ${\overline \Lambda}(\infty)=0.83750$ is indicated as a black horizontal line.
}
\label{sc-var}
\end{figure}

\begin{figure}
    \centerline{\includegraphics[width=3in,keepaspectratio,clip=]{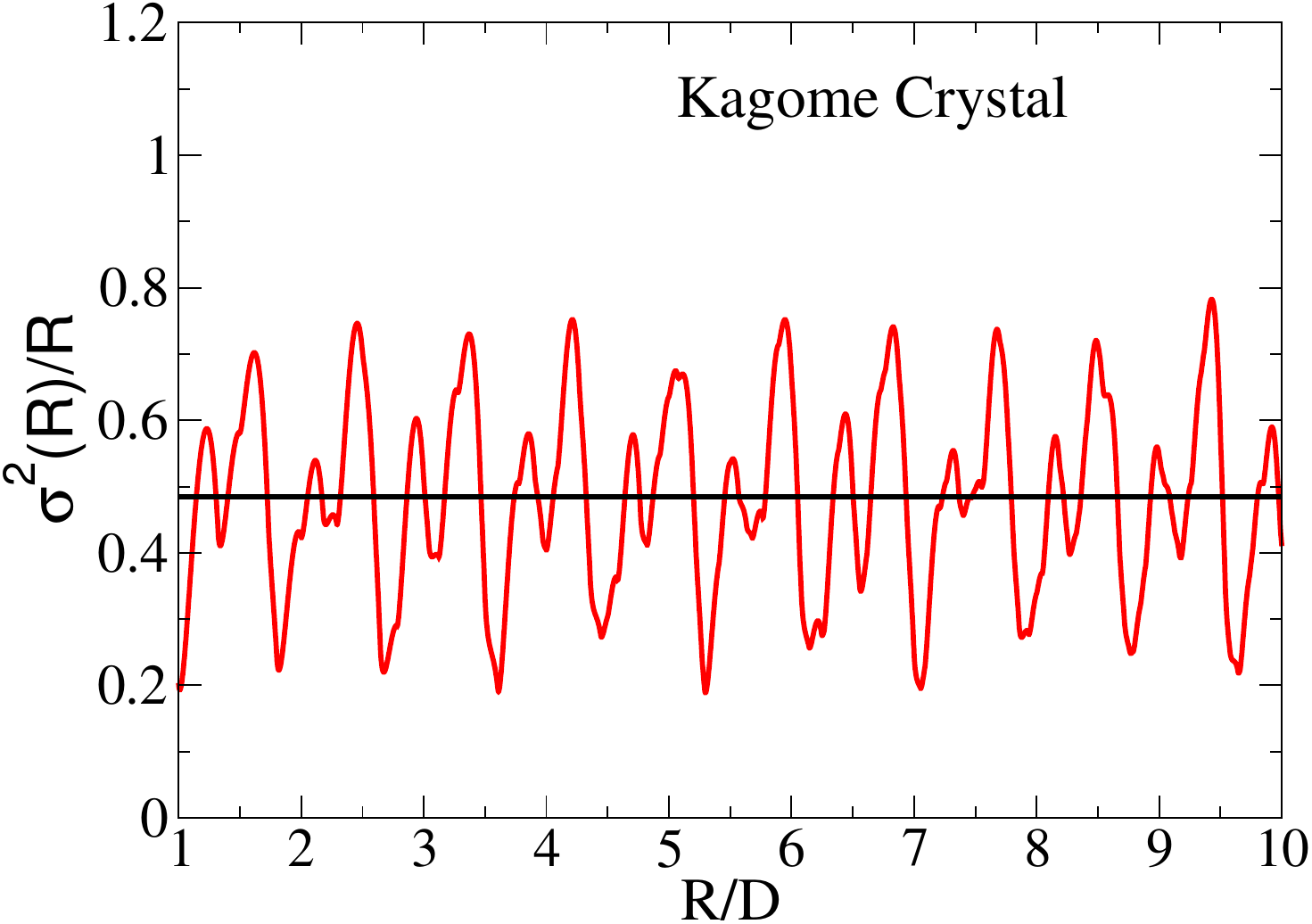}}
    \caption{The scaled local variance $\sigma_N^2(R)/R$ versus $R/D$ for the kagome crystal, where $D$ is the nearest-neighbor distance and we take $\phi=3\pi/(8\sqrt{3})\approx 0.6802$. The average ${\overline \Lambda}(\infty)=0.48411\ldots$ is indicated as a black horizontal line.
}
\label{kagome-var}
\end{figure}

For Bravais lattices in $\mathbb{R}^d$, it is well known that
the scaled variance $\sigma_N^2(R)/R^{d-1}$ oscillates around a well-defined average value ${\overline \Lambda}(\infty)\equiv \lim_{L \to \infty}{\overline \Lambda}(L)$
with a period of the order of $D/2$, where $D$ is the nearest neighbor distance .
This function appears to be chaotic [see, {\it e.g.}, Fig.~\ref{sc-var} and~\ref{kagome-var}], but is determined by a certain sum involving the square of Bessel functions
of order $d/2$ and hence is purely deterministic~\cite{To03a}. An illustration of this situation is given in Ref. \cite{To03a}, where $\sigma_N^2(R)/R^2$ for the square lattice for small values of $R$ is presented.
Here we will analyze the small-$R$ behavior of the scaled variance and its corresponding cumulative moving average ${\overline \Lambda}(L)$ for the simple cubic lattice. 

First, note that for any single, infinitely large periodic point configuration in $d$-dimensional Euclidean space
$\mathbb{R}^d$ at number density $\rho$, the radial (angular-averaged) pair correlation function can be written as~\cite{To18a}
\begin{equation}
g_2(r)=  \sum_{i=1}^{\infty}\frac{Z_i}{\rho s_1(r_i)}\;\delta(r-r_i),
\label{g2-period}
\end{equation}
where $Z_i$ is the expected coordination number at radial distance $r_i$
( {\it i.e.}, the  expected number of points that are exactly at a distance $r=r_i$ from a point of the point process) such that $r_{i+1} > r_i$ and $\delta(r)$ is a radial Dirac delta function.
Second, we can re-write relation Eq.~\eqref{local-1} for the local number variance as
\begin{equation}
\sigma_N^2(R)=
 \rho v_1(R)\Big[ 1- \rho v_1(R)+ \rho\int_{\mathbb{R}^d} g_2({\bf r})
\alpha_2(r;R) d{\bf r}\Big].
\label{local-3}
\end{equation}
Substitution of Eq.~\eqref{g2-period} into Eq.~\eqref{local-3} yields
\begin{equation}
\sigma_N^2(R) =
  2^d \phi \left(\frac{R}{D}\right)^d \Big[ 1-  2^d \phi \left(\frac{R}{D}\right)^d + \sum_{j=1}^\infty Z _j  \alpha_2(r _j ;R)  \Big]
 \label{local-4}
\end{equation}
where $D$ is the nearest-neighbor distance. Note that this relation is identical to Eq. (78) in Ref.~\cite{To03a}, which was derived differently, namely, via the {\it volume-average} formulation.
Systems with Bragg peak diffraction, such as periodic point patterns, can display small-scale fluctuations of the order of the nearest-neighbor separation distances~\cite{To03a}, even if hyperuniform.

\begin{figure}
    \centerline{\includegraphics[width=3in,keepaspectratio,clip=]{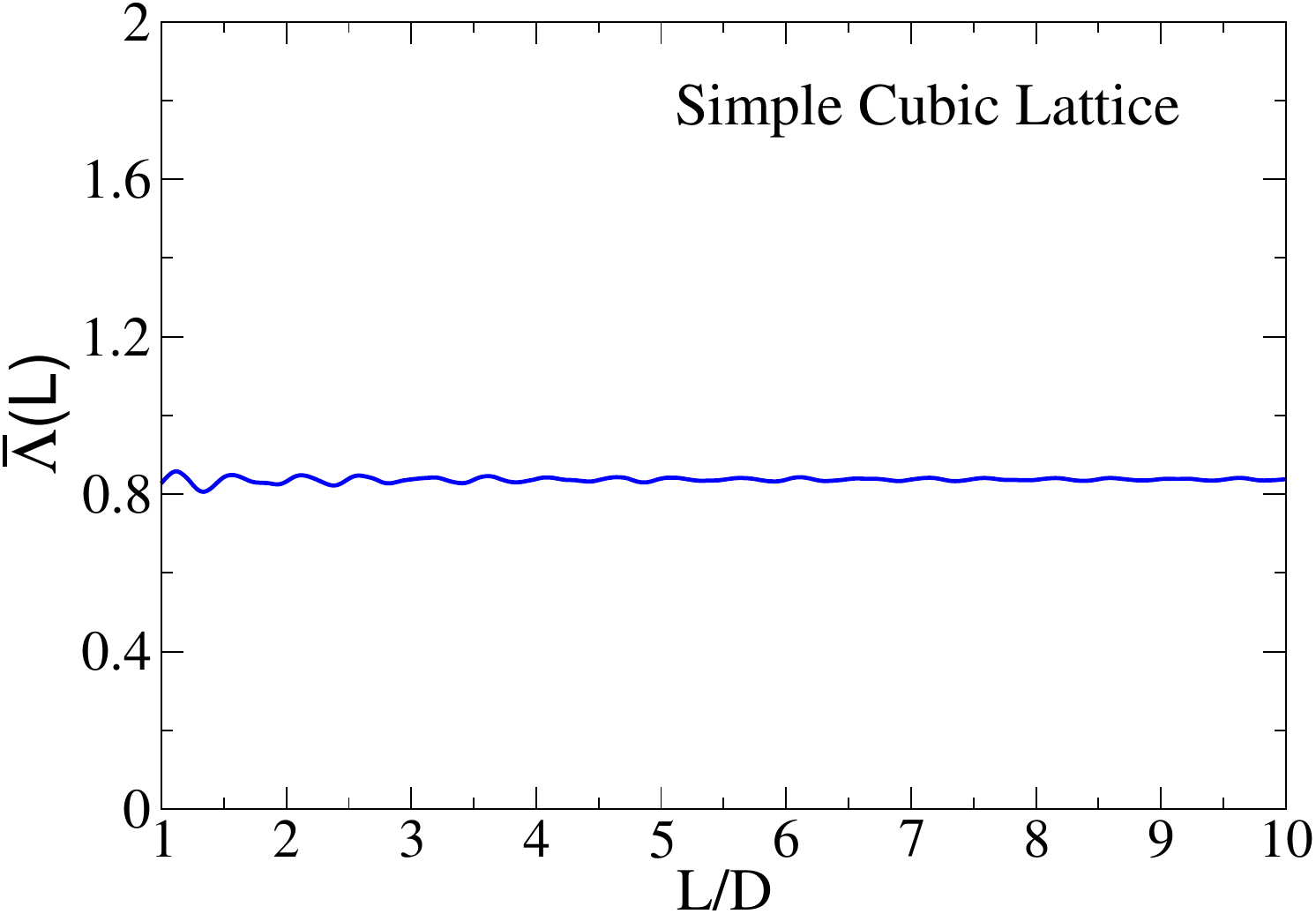}}
    \caption{ ${\overline \Lambda}(L)$ versus $L/D$ for the simple cubic lattice, where $D$ is the nearest-neighbor distance,  and we take $\phi=\pi/6\approx 0.5236$.
}
\label{sc-Lambda}
\end{figure}

For the simple cubic lattice, one can extract the coordination numbers $Z_+,Z_2, \ldots$ from the theta series~\cite{Co93}, which,  when substituted into Eq.~\eqref{local-4},  yields the corresponding local number variance.
Figure~\ref{sc-var} shows the scaled variance as a function of $R/D$ as well as its infinite-distance average ${\overline \Lambda}(\infty)=0.83750$~\cite{To03a}. The corresponding plot of the cumulative moving average ${\overline \Lambda}(L)$ is depicted in Fig.~\ref{sc-Lambda}, which is seen to settle to its infinite-distance value rather rapidly (several lattice spacings), but not as rapidly as the disordered class I hyperuniform systems examined here.

\begin{figure}
     \centerline{\includegraphics[width=3in,keepaspectratio,clip=]{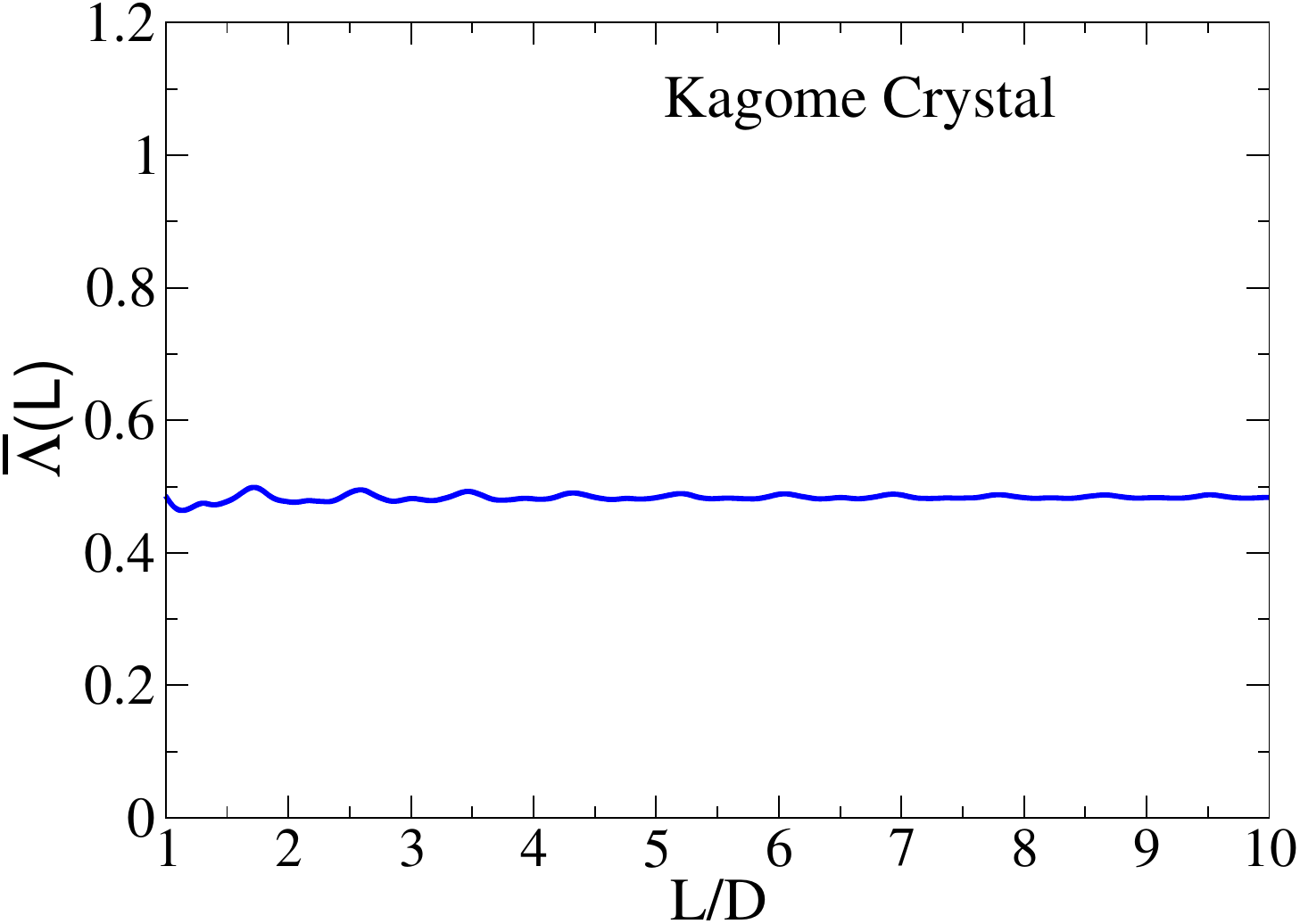}}
    \caption{ ${\overline \Lambda}(L)$ versus $L/D$ for the kagome crystal, where $D$ is the nearest-neighbor distance and we take $\phi=3\pi/(8\sqrt{3})\approx 0.6802$.}
\label{kagome-Lambda}
\end{figure}

For non-Bravais lattices with crystallographic symmetry, it is expected that ${\overline \Lambda}(L)$ settles more rapidly to its infinite-distance value as the particle basis increases.
Here we examine the kagome crystal, which should be noted to have only a 3-particle basis.
Figure~ \ref{kagome-var} shows the scaled variance as a function of $R/D$ as well as its infinite-distance average ${\overline \Lambda}(\infty)=0.48411\ldots$~ \cite{To03a}. 
The corresponding plot of the cumulative moving average ${\overline \Lambda}(L)$ is depicted in Fig.~\ref{kagome-Lambda}, which is seen to settle to its infinite-distance value rather rapidly (few lattice spacings).

\subsection{Quasicrystals}
\label{sec:quasicrystal}

Unlike crystals, quasicrystals are characterized by the absence of translational periodicity, but possess quasiperiodic translational order. On the other hand, they possess long-range orientational order but with prohibited crystallographic symmetries~\cite{Le84,Lev86,Sh84}. The structure factor of a quasicrystal displays a dense set of Bragg peaks. These systems, therefore, cannot be characterized by the behavior of the structure factor near the origin as in Eq.~\eqref{eq:Skalpha}. However, as shown in Ref.~\cite{Og17}, Eq.~\eqref{eq:sigma_F_space} can be alternatively written as
\begin{equation}
    \sigma_N^2(R) = -\rho\, v_1(R) \left[ \frac{1}{(2\pi)^d} \int_0^{\infty} Z(k) \, \frac{\partial \tilde{\alpha}_2(k; R)}{\partial k} \, dk \right],
\end{equation}
where
\begin{equation}
    Z(k) = \int_0^k S(q)\, s_d\, q^{d-1} \, dq,
\end{equation}
and $s_d = d \, \pi^{d/2}/\Gamma(1+d/2)$ is the surface area of a $d$-dimensional sphere with unit radius.
This definition has the advantage that the cumulative intensity function $Z(k)$ is smoother than $S(k)$, and allows for to extraction of the value of $\alpha$ as $Z(k) \sim k^{\alpha + 1}$ as $k\to 0$. In the cases in which $S(k)$ is a smooth function, this definition coincides with Eq.~\eqref{eq:Skalpha}.

\begin{figure}
    \centering
    \includegraphics[width=3in]{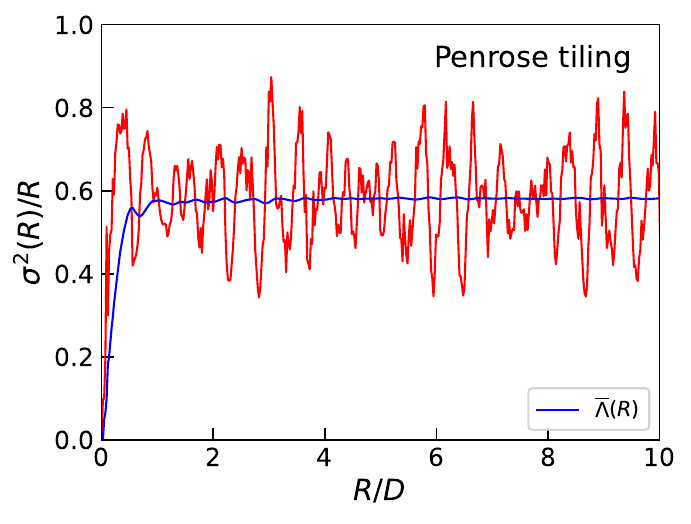}
    \caption{Scaled number variance (red) and its cumulative moving average (blue) for the Penrose tiling quasicrystal.}
    \label{fig:num_var_Penrose}
\end{figure}

Here, we focus on quasicrystals belonging to class I. To be specific, we will consider the points obtained from the vertices of the obtuse rhombi that compose the Penrose tiling~\cite{Za09,Li17a}.
In this case, we numerically compute the local number variance, and we show in Fig.~\ref{fig:num_var_Penrose} the result we obtain. 
Similarly to what we observed for the lattices, the local number variance of the Penrose quasicrystal shows undamped oscillations around the average value (already observed in Ref.~\cite{Li17a,Og17}), and the cumulative moving average $\Lambda(R)$ saturates to the asymptotic value over a few average particle distances, hence implying that quasicrystals are uniform from short to large length scales.

\subsection{$g_2$-invariant unit step function}
\label{sec:g2invariant}

A $g_2$-invariant process is obtained by choosing a nonnegative pair correlation function $g_2$ that remains invariant over a nonvanishing density range, while all other macroscopic variables remain fixed~\cite{To02b}. Such processes have a critical density above which the nonnegativity of the structure factor would be violated. Therefore, at the critical point, if $S(k=0)=0$ and the system is realizable, it is hyperuniform.

In Ref.~\cite{To03a}, the simplest $g_2$-invariant process has been identified as the one in which the radial distribution function is defined by the unit step function \cite{To02c}, {\it i.e.}, 
\begin{equation}
g_2(r)=\Theta(r-D)=\left\{\begin{array}{ll}
0, \qquad & r\leq D, \nonumber \\
1, \qquad & r>D,\\
\end{array}
\right.
\label{step}
\end{equation}
The corresponding structure factor for packing fraction $\phi=\rho v_1(D/2)$ in the range $0 \le \phi \le \phi_c$, where $\phi_c=1/2^d$, is known~\cite{To03a} and given by
\begin{equation}
S(k)=1-\Gamma(1+d/2) 
\left(\frac{2}{kD}\right)^{d/2}
\left(\frac{\phi}{\phi_c}\right) J_{d/2}(kD).
\end{equation}
The maximal or critical value $\phi=\phi_c$ corresponds to a hyperuniform state of class I.

For $d=2$, we derive the expression for the scaled number variance by using Eq~\eqref{step} in Eq.~\eqref{local-1} and setting $\phi = \phi_c$ and $D=1$ (thus with $\rho = 1/\pi$), getting
\begin{equation}
    \sigma_N^2(R) = \frac{\left(2 R^2+1\right) \sqrt{4 R^2-1}}{4 \pi}-\frac{2}{\pi} R^2 \left(R^2-1\right) \csc ^{-1}(2 R),
\end{equation}
which, in the large-$R$ limit, gives
\begin{equation}
    \Sigma^2(R) = 1 - \frac{1}{40 R^2} + O \left(\frac{1}{R^4}\right).
\end{equation}
The same procedure in 3D gives instead (with  $\rho = 3/4\pi$)
\begin{equation}
\sigma_N^2(R)= \frac{9}{16} R^2 -\frac{1}{32}, \quad R>D/2,
\end{equation}
which is equivalent to 
\begin{equation}
    \Sigma^2(R) = 1- \frac{1}{18 R^2}
\end{equation}
\begin{figure}
    \centering
    \includegraphics[width=3in]{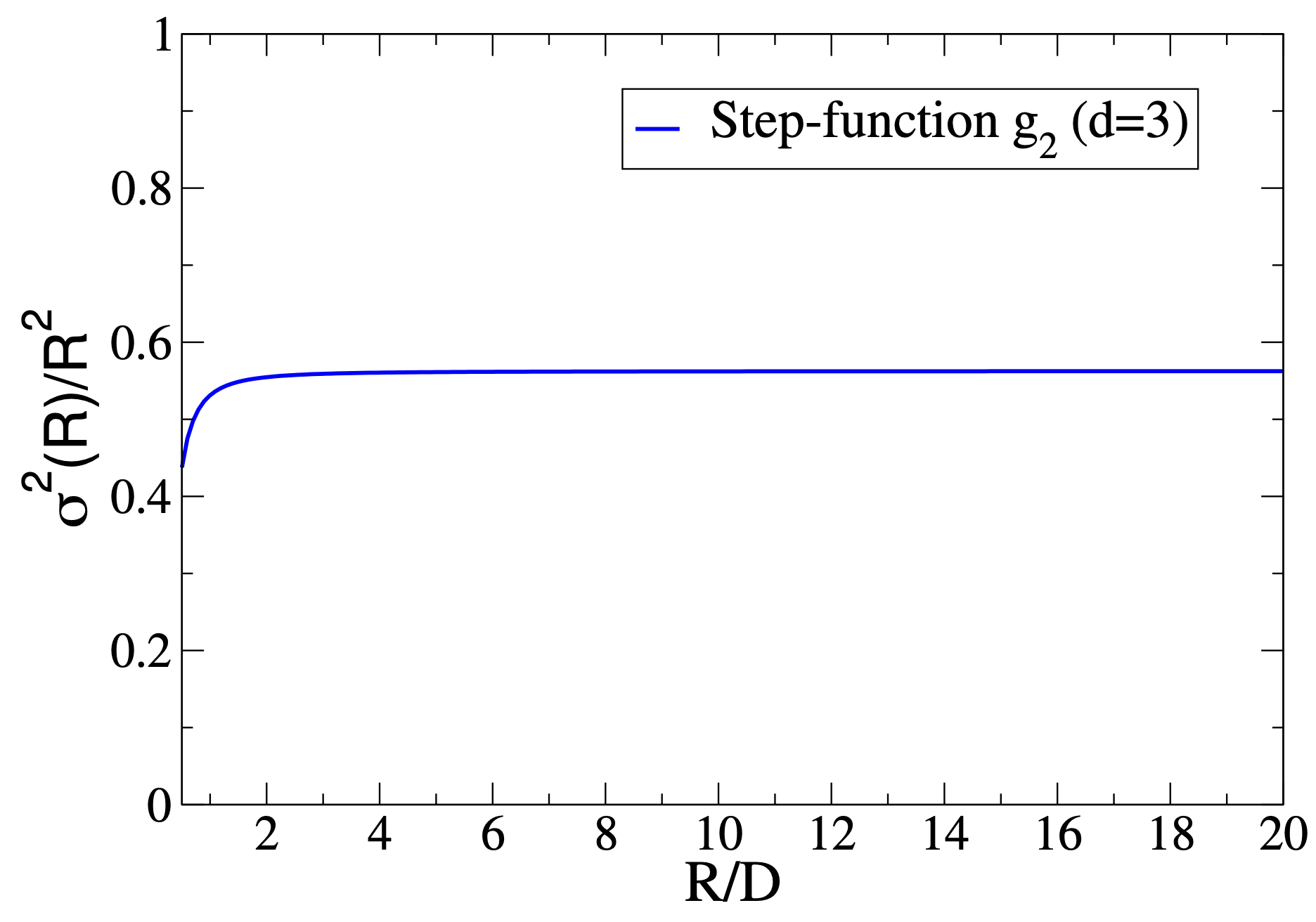}
    \caption{Scaled variance $\sigma_N^2(R)/R^2$ versus $R$ for the unit-step-function $g_2$ at the hyperuniform state in three dimensions.}
    \label{step-2}
\end{figure}

Figure \ref{step-2} shows the scaled variance $\sigma_N^2(R)/R^2$ as function
of the window radius $R$ at the hyperuniform state in three dimensions.
It is seen that $\sigma_N^2(R)/R^2$ achieves its asymptotic value for small $R/D$,   {\it i.e.} , when it is of the order of the mean-nearest neighbor distance $R/D \sim 3.22$. 

\vspace{4mm}

\subsection{Stealthy Hyperuniform Disordered Ground States}
\label{sec:SHU}

Stealthy hyperuniform (SHU) systems belong to Class I, in that their structure factor vanishes not only at $k=0$, but rather for a whole range of wave numbers around the origin up to a maximal value $K$, usually denoted as the exclusion region
\begin{equation}
    S({\bf k}) = 0, \quad 0\leq |{\bf k}|<K.
\end{equation}
In some sense, stealthy systems can be intended as the limit $\alpha \to \infty$. Perfect crystals are SHU systems in which $K = |{\bf k}_{\text{Bragg}}|$, {\it i.e.}, the position of the first Bragg peak. 
More importantly, SHU systems can also be disordered, thus being isotropic and without Bragg peaks. 
At the same time, they display characteristics that are typical of crystals, such as the absence of arbitrarily large holes in the infinite system-size limit and the absence of single scattering from intermediate to infinite wavelengths.

The size of the exclusion region $K$ characterizes the properties of disordered SHU systems, and in particular, for larger $K$, more wave vectors are constrained. It is then common to introduce the relative fraction of independently constrained wave vectors, denoted as the stealthiness parameter $\chi = M(K)/[d(N-1)]$~\cite{To15}, that in the thermodynamic limit reduces to
\begin{equation}
    \chi = \frac{v_1(K)}{2\, \rho \, d \, (2\pi)^d}. 
\end{equation}
For small values of $0 \leq \chi < 1/2$, SHU systems are disordered~\cite{To15}, while transitions to a crystal phase can occur at $\chi = 1/2$, depending on the dimensionality $d$. In this Section, we will focus on the disordered regime $\chi < 1/2$.

\subsubsection{Small $\chi$}

\begin{figure}
    \centering
    \includegraphics[width=3in]{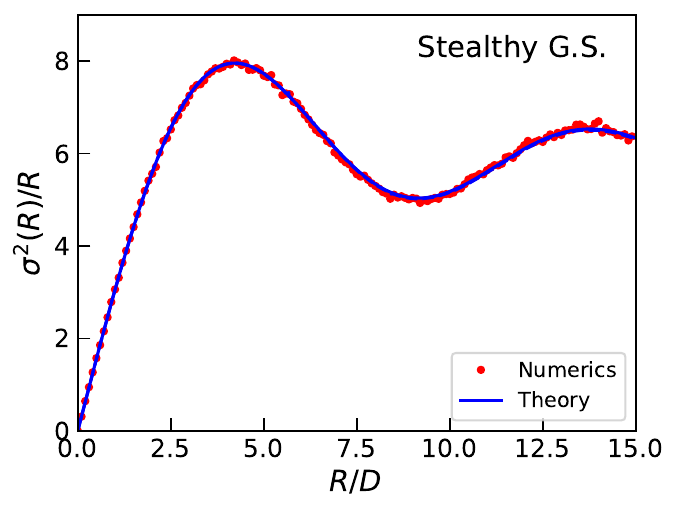}
    \caption{Scaled number variance for stealthy hyperuniform ground states in 2D with $\chi = 0.0025$. The red points are obtained numerically for a finite-size system with $N=4000$ and averaging over 1000 realizations. The blue dashed and solid lines are the analytical results with and without the $O(\chi)$ correction in the structure factor, and they both agree very well with the numerical data.}
    \label{fig:num_var_SHU}
\end{figure}

Let us consider the limit in which $\chi \ll 1/2$. Following Refs.~\cite{To15,Morse2024Ordered}, we express the total correlation function as
\begin{equation}
    \rho \tilde{h}(k) = -1 + \Theta(k-K)[1+ d \, \alpha(k;K) \, \chi + O(\chi^2)],
\end{equation}
so that the structure factor is expressed as
\begin{equation}
    S(k) = \Theta(k-K)[1+ d \, \alpha(k;K) \, \chi + O(\chi^2)].
\end{equation}
Let us consider the leading order contribution in $\chi$, that is $S(k) = \Theta(k-K)[1 + O(\chi)]$. Using Eq.~\eqref{eq:sigma_F_space}, we get
\begin{widetext}
    \begin{equation}
    \label{eq:small-chi}
    \sigma_N^2(R) = \frac{2^{-d} \pi ^{d/2} R^d \left(2^d-\Gamma (d+1) (R K)^d \, _2\tilde{F}_3\left(\frac{d+1}{2},\frac{d}{2};\frac{d+2}{2},\frac{d+2}{2},d+1;-R^2 K^2\right)\right)}{\Gamma
   \left(\frac{d}{2}+1\right)} + O(\chi).
\end{equation}
\end{widetext}
In 2D, the expression Eq.~\eqref{eq:small-chi} gives
\begin{equation}
\label{eq:SHU_2d}
    \Sigma^2(R) = 1-\frac{\cos(2KR)}{2KR} + O\left(\frac{1}{R^2}\right) + O(\chi),
\end{equation}
while in 3D it reduces to
 \begin{equation}
     \Sigma ^2(R) =  1 - \frac{\sin (2 KR)}{ 2 KR}  + O\left(\frac{1}{R^2}\right) + O(\chi). 
 \end{equation}
We report in Fig.~\ref{fig:num_var_SHU} the comparison between the analytical prediction and the numerical results, which show excellent agreement.

\subsubsection{Large $\chi$}

The analytical derivation presented in the previous section is valid only for small values of $\chi$. 
It is interesting, however, to investigate what happens at large values of $\chi \sim 0.5$. 
To do so, we numerically compute the scaled number variance for stealthy disordered configurations obtained through the collective-coordinate optimization procedure (CCO) with the addition of soft-core repulsion, as discussed in recent works~\cite{Kim_2025_DenseSphere,Vanoni2025Dynamical}.
We report the results for $\sigma_N^2(R)/R$ in a 2D stealthy system having $\chi = 0.45$ and volume fraction of the soft-core repulsion $\phi = 0.68$.

\begin{figure}
    \centering
      \includegraphics[width=3in]{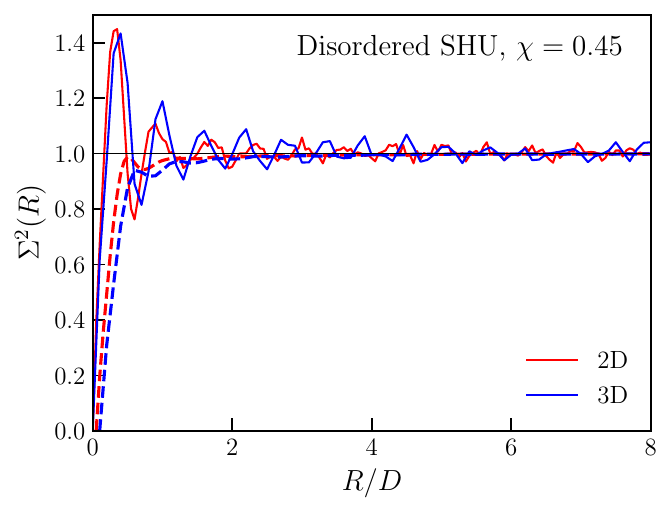}
     \caption{Scaled number variance (solid lines) and cumulative moving average (dashed lines) for a disordered stealthy hyperuniform system with $\chi = 0.45$ in 2D (red) and 3D (blue).}
\label{fig:num_var_stealthy-highChi}
\end{figure}

We can now compare the results of the local number variance of lattices and disordered SHU systems. 
Let us first note that, at small $\chi$ values, the number variance attains its asymptotic scaling at large values of $R/D$. This is consistent with the small number of constrained wave vectors. 
In fact, uniformity is expected to be recovered at scales $R \sim 1/K$.

By increasing $K$, and thus $\chi$, the number of constrained wave vectors increases, and local uniformity is recovered at smaller length scales. For $\chi \lesssim 1/2$ [see Fig.~\ref{fig:num_var_stealthy-highChi}], the cumulative moving average~\eqref{eq:Lambda} saturates to its asymptotic value at scales compatible to the ones observed in lattices [cf. Fig.~\ref{sc-Lambda} and~\ref{kagome-Lambda}]. Indeed, lattices are recovered for $\chi = 1/2$. However, as long as $\chi<1/2$, the configurations are disordered, explaining why the short-scale oscillations present in $\Sigma^2(R)$ in Fig.~\ref{fig:num_var_stealthy-highChi} are damped at larger scales. In the case of lattices, instead, such oscillations are undamped due to the order and symmetries of the system~\cite{To03a}.

 \subsection{One-component plasma}
\label{sec:onecompplasma}

A one-component plasma is a system of identical point particles with charge $e$ at equilibrium, interacting via the log-Coulomb potential and in a uniform background of opposite charge that enforces the overall neutrality. 
The total potential energy of the system is given by
\begin{equation}
    \Phi_N({\bf r}^N) = N \sum_{i=1}^N V({\bf r}_i) - \sum_{i<j}^N \ln(|{\bf r}_i - {\bf r}_j|),
\end{equation}
where $V({\bf r})$ is the background potential.

\subsubsection{Two dimensions}

The one-component plasma is a point process that is naturally embedded in two dimensions, as the Coulomb potential in 2D is log-Coulomb.
The total correlation function for the one-component plasma (Ginibre ensemble) is known~\cite{Ja81} to be
\begin{equation}
    h(r) = - \exp(-\rho \pi r^2),\qquad \tilde{h}(k) = -\frac{1}{\rho}\exp \left(-\frac{ k^2}{4 \pi \rho}\right)
\end{equation}
and the corresponding structure factor is
\begin{equation}
    S(k) = 1 - \exp \left(-\frac{ k^2}{4 \pi \rho}\right) .
\end{equation}
One thus immediately finds the number variance
\begin{equation}
    \sigma_N^2(R) = \rho \pi R^2 e^{-2 \pi  \rho  R^2} \left(I_0\left(2 \pi  R^2 \rho \right)+I_1\left(2 \pi R^2 \rho \right)\right).
\end{equation}
In the large-$R$ limit, the asymptotic behavior of the scaled structure factor is given by
\begin{equation}
    \Sigma^2(R) = 1 - \frac{1}{16 \pi  \rho  R^2} -\frac{3}{512 \pi ^2 \rho ^{2} R^4} + O\left( \frac{1}{R^6} \right),
\end{equation}
and thus, the asymptotic value is reached with $1/R^2$ corrections.

\begin{figure}
    \centering
    \includegraphics[width=3in]{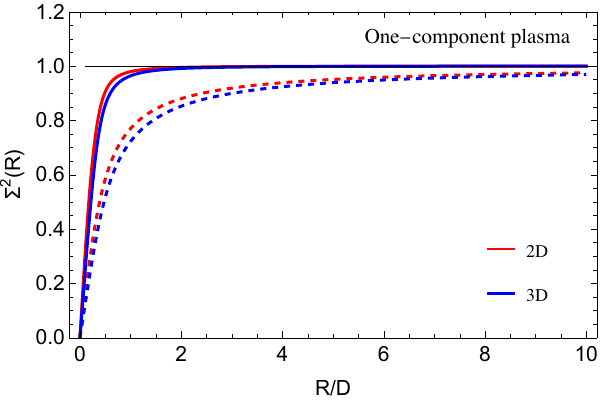}
    \caption{ Solid lines represent the scaled number variance for the one-component plasma in 2D (blue) and the generalization to 3D (red). The dashed lines, following the same color scheme as for the scaled number variance, represent the cumulative moving average $\overline{\Lambda}(R)/\overline{\Lambda}$ .}
    \label{fig:1CP}
\end{figure}

\subsubsection{Generalization to three dimensions}
Wang and Torquato~\cite{Wang2024Designer} showed the existence of a point process that generalizes the form of the OCP pair correlation to three dimensions.
Specifically, this correlation function is given by
\begin{equation}
    g_2(r) = 1 - \exp\left( - \frac{4\pi}{3} r^3\right).
\end{equation}
The corresponding structure factor is given by
\cite{Wang2024Designer}

 \begin{align}
    S(k) =& 1 - {}_1F_4\left(1; \frac{1}{3}, \frac{2}{3}, \frac{5}{6}, \frac{7}{6}; -\frac{k^6}{20736\pi^2} \right) \\
    &+ \frac{2\pi}{3\sqrt{3}} \left[\operatorname{ber}_{\frac{2}{3}} \left(\frac{(k^2)^{3/4}}{3\sqrt{\pi}} \right) + \operatorname{ber}_{-\frac{2}{3}} \left(\frac{(k^2)^{3/4}}{3\sqrt{\pi}} \right)\right] \nonumber ,
 \end{align}
where $_pF_q$ is the generalized hypergeometric function and $\operatorname{ber}_{\nu}(x)$ is the real Kelvin function.

With the usual method, we can compute the number variance, which results in
\begin{align}
    \sigma_N^2(R) =& -\frac{64}{3} \pi ^2 R^6 E_{-\frac{1}{3}}\left(\frac{32 \pi  R^3}{3}\right)\nonumber\\
    &+e^{-\frac{32 \pi  R^3}{3}} \left(2 \pi  R^3+\frac{1}{16}\right)\nonumber\\
    &+\left(\frac{\pi }{6}\right)^{2/3} R^2
   \Gamma \left(\frac{1}{3}\right)-\frac{1}{16},
\end{align}
being $E_{a}(x)$ the exponential integral function. The scaled number variance is, in the large $R$ limit,
\begin{equation}
    \Sigma^2(R) = 1 -\frac{1}{16 \left(\frac{\pi }{6}\right)^{2/3} \Gamma \left(\frac{1}{3}\right) R^2}+O\left(\frac{1}{R^5}\right),
\end{equation}
and therefore, analogously to the two-dimensional case, the asymptotic value is reached with $1/R^2$ corrections. The comparison between the two- and three-dimensional results is reported in Fig.~\ref{fig:1CP}.

\subsection{Gaussian $g_2$}
\label{sec:gaussian}

\begin{figure}
    \centering
    \includegraphics[width=3in]{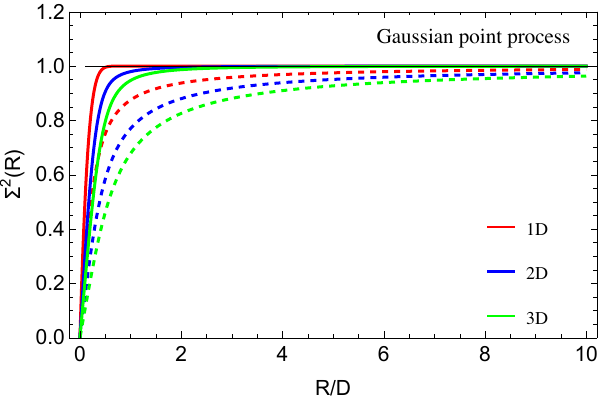}
    \caption{Solid lines represent the scaled number variance for the Gaussian point process in 1D (red), 2D (blue), and 3D (green). Following the same color scheme as for the scaled number variance, the dashed lines report the cumulative moving average $\overline{\Lambda}(R)/\overline{\Lambda}$.}
    \label{fig:gaussian}
\end{figure}

Following Wang and Torquato~\cite{Wang2024Designer}, we consider a hyperuniform system whose correlation function is Gaussian
\begin{equation}
    g_2(r) = 1 - \exp\left( -\pi r^2 \right),
\end{equation}
which is a legitimate choice in $d=1,\,2, \, 3$.
It has been shown~\cite{St76,Louis2000Meanfield,Za08} that such correlation functions are realizable by point processes belonging to class I hyperuniformity with $\alpha = 2$~\cite{Wang2024Designer}. 
In 2D, it coincides with the one-component plasma $g_2$. The analytical expressions for $\sigma_N^2(R)$ are
\begin{equation}
    \sigma_N^2(R) = \frac{-2 \pi  R \, \text{erf}\left(2 \sqrt{\pi } R\right)-e^{-4 \pi  R^2}+2 \pi  R+1}{\pi }, \quad d=1,
\end{equation}
\begin{equation}
    \sigma_N^2(R) = \pi  e^{-2 \pi  R^2} R^2   \left[ I_0\left(2 \pi  R^2\right)+I_1\left(2 \pi  R^2\right)  \right] , \quad d=2,
\end{equation}
\begin{align}
    \sigma_N^2(R) =& 2R^2 + \frac{4\pi}{3}R^3 \text{erfc}\left(2 \sqrt{\pi } R\right) \nonumber \\
    &+ \frac{e^{-4 \pi  R^2} \left(1-2 \pi  R^2\right)-1}{3 \pi }, \quad d=3.
\end{align}
For large $R$
\begin{equation}
    \Sigma^2(R) \simeq 
    \begin{cases}
        1 - \exp(-4\pi R^2)/(8\pi R^2), \quad & d=1,\\
        1 - 1/(16 \pi R^2), \quad & d=2,\\
        1 - 1/(6\pi R^2), \quad & d=3.\\        
    \end{cases}
\end{equation}
The comparison between the results in one, two, and three dimensions is reported in Fig.~\ref{fig:gaussian}.

\section{Results -- Class II}
\label{sec:classII}

Class II hyperuniform systems are distinguished by a number variance that scales as $\sigma_N^2(R) \sim R^{d-1} \ln (R) + O(R^{d-1})$, see Eq.~\eqref{eq:classes}. As a consequence, the approach to the asymptotic scaling is attained with a slow, logarithmic correction, namely $\Sigma^2(R)- 1 \propto 1/\ln(R)$. For this reason, as we will now discuss for the specific example of ‘Fermi-sphere’ point processes, class II hyperuniform systems provide the weakest form of uniformity across length scales among hyperuniform systems.

\subsection{‘Fermi-sphere’ point processes}
\label{sec:Fermi}

The Fermi-sphere point process~\cite{To08b,Sc09} constitutes a $d$-dimensional generalization of one-dimensional determinantal point processes associated with the Gaussian Unitary Ensemble (GUE) eigenvalue statistics, the nontrivial zeros of the Riemann zeta function, and the spatial distribution of an ideal one-dimensional Fermi gas. This disordered hyperuniform process, classified as type II, exactly corresponds to the ground-state configuration of a system of noninteracting spin-polarized fermions with a completely occupied Fermi sea in $d$-dimensional Euclidean space.
The 2-point function is known analytically in any dimension, and it is given by
\begin{equation}
\label{eq:g2Fermi}
    g_2(r) = 1-2^d \Gamma\left( 1+d/2 \right)^2 \frac{J_{d/2}^2(Kr)}{(Kr)^d}
\end{equation}
with
\begin{equation}
    K = \rho^{1/d} 2 \sqrt{\pi} \Gamma\left( 1+d/2 \right)^{1/d}.
\end{equation}

Let us consider $d=1,\,2,\,3$. In 1D, use of Eq.~\eqref{eq:g2Fermi} into Eq.~\eqref{local-1} gives
\begin{align}
    \sigma_N^2(R) =&  \frac{-\text{Ci}(4 \pi  R)-4 \pi  R \, \text{Si}(4 \pi  R)+2 \pi ^2 R}{\pi^2}  \nonumber\\
    &+\frac{\ln (4 \pi  R)-\cos (4 \pi  R)+\gamma +1}{\pi ^2},
\end{align}
which, in terms of the scaled number variance $\Sigma^2(R)$ in the large $R$ limit, is
\begin{equation}
    \Sigma^2(R) \simeq 1+ \frac{\gamma +1+\ln (\pi )+2 \ln (2)}{\ln(R)}.
\end{equation}
We can notice that the first correction to $\Sigma^2(R)$ is logarithmic in $R$, signaling a much slower convergence to the asymptotic scaling, and therefore reduced local uniformity.

In 2D, a similar analysis gives
\begin{align}
    \sigma_N^2(R) =& \pi  R^2   \left[ \pi  R^2 \, _2F_3\left(\frac{3}{2},\frac{3}{2};2,3,3;-16 \pi  R^2\right)\right.\\
    &+\left.\, _2F_3\left(\frac{1}{2},\frac{1}{2};1,1,2;-16 \pi  R^2\right)  \right]
 \end{align}
leading to a scaled number variance that in the large-$R$ limit scales as
\begin{equation}
    \Sigma^2(R)\simeq  1 + \frac{2 \gamma -4+\ln (\pi )+ \ln (4096)}{2 \ln (R)}.
\end{equation}
We can observe that, also in 2D, the leading correction is logarithmic.
  \begin{figure}[t!]
    \centering
    \includegraphics[width=3in]{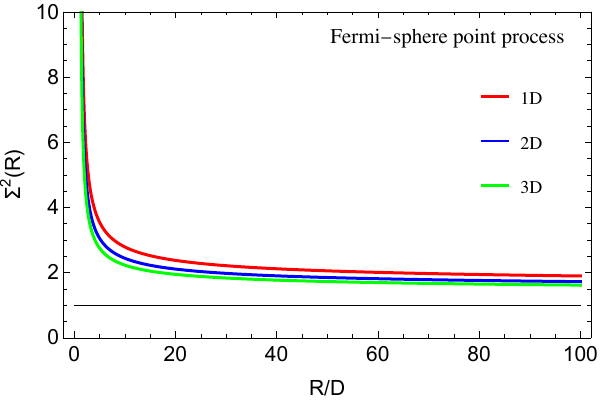}
    \caption{Scaled number variance for the ‘Fermi-sphere’ point process in 1D (blue), 2D (orange), and 3D (green). We have scaled the curves so that they all asymptote to 1 for better comparison.}
    \label{fig:sigma_DPP}
\end{figure}
 In 3D, the number variance can be computed exactly using the usual procedure. We report here the large-$R$ behavior of the scaled number variance, given by
\begin{equation}
\label{eq:sigma_DPP_corr}
    \Sigma^2(R) \simeq  1 + \frac{-3+6 \gamma +14 \ln (2)+\ln (9)+4 \ln (\pi )}{6 \ln (R)} \nonumber,
\end{equation}
displaying a logarithmic correction to the asymptotic scaling. We report the direct comparison between the results obtained for $d=1,~2$ and $3$ in Fig.~\ref{fig:sigma_DPP}.

In agreement with the general discussion we presented at the beginning of this section, ‘Fermi-sphere’ point processes display logarithmic corrections to the asymptotic hyperuniform scaling. Consequently, class II hyperuniform systems present the weakest form of uniformity at small and intermediate length scales among all hyperuniformity classes.

\section{Results -- Class III}
\label{sec:classIII}

In this section, we address an exactly solvable model belonging to class III hyperuniformity, namely $d$-dimensional class III perturbed lattices, first studied by Kim and Torquato in Ref.~\cite{Ki18a}. 

\subsection{Perturbed points configurations}
\label{sec:perturbed}

Perturbed point configurations generated from a hyperuniform system are obtained by considering a hyperuniform point configuration in which particle $i$ is at position ${\bf r}_i$, and its position is perturbed to be ${\bf r}_i + {\bf u}({\bf r}_i)$. We assume that the stochastic displacement vectors ${\bf u}({\bf r})$ are distributed according to an identical and isotropic singlet probability density function $f_1({\bf u})$.
In this Section, we will consider the specific case of class III perturbed lattices in $\mathbb{R}^d$, for which we take, following Ref.~\cite{Ki18a}, the singlet density function to be
\begin{equation}
    f_1(\mathbf{r};\delta,\alpha) \equiv 
    \begin{cases}
        K(d,\alpha,\delta), & |{\bf r}|\leq \delta\\
        K(d,\alpha,\delta)\left(\frac{|{\bf r}|}{\delta}\right)^{-d-\alpha}, & |{\bf r}| > \delta.
    \end{cases}
\end{equation}
The normalization constant is $K(d,\alpha,\delta) = \Gamma(1+d/2)\alpha/[\pi^{d/2}(d + \alpha)\delta^d]$. The parameter $\alpha \in (0,1)$ determines the hyperuniformity exponent, while a typical length scale is set by $\delta \in (0,\infty)$.

If the displacements of distinct particles are uncorrelated, the structure factor can be expressed as~\cite{Ga04,Ga04b}
\begin{equation}
    S(\mathbf{k}) = (1 - |\tilde{f}_1(\mathbf{k})|^2) + |\tilde{f}_1(\mathbf{k})|^2 S_0(\mathbf{k}),
\end{equation}
where the characteristic function $\tilde{f}_1(\mathbf{k})$ is the Fourier transform of $f_1({\bf r})$ and $S_0(\mathbf{k})$ is the structure factor of the initial configuration.

We are interested in the large $R$ behavior, and therefore we can consider the small-$k$ expansion of the characteristic function $\tilde{f}_1(\mathbf{k})$, whose complete form can be found in Ref.~\cite{Ki18a}. Also, since the original configuration is a lattice, we can set $S_0(k) = 0$ for $k<O(1)$, corresponding to the position of the first Bragg peak. We get
\begin{equation}
    \tilde{f}_1(k;\delta,\alpha) = 1 - A(d,\alpha) \left( \frac{k \delta}{2} \right)^{\alpha} + O\left(k^2\right).
\end{equation}
At this point, it is straightforward to compute the structure factor
\begin{equation}
    S(k) = 2 A(d,\alpha) \left( \frac{k \delta}{2} \right)^{\alpha} - A(d,\alpha)^2 \left( \frac{k \delta}{2} \right)^{2 \alpha}
    + O(k^2) ,
\end{equation}
and we can use Eq.~\eqref{eq:sigma_F_space} to compute the number variance.
\begin{widetext}
In $d=2$ and for $\alpha < 1/2$, after setting $\rho = 1/(\pi \delta^2)$ for convenience, the number variance is given by
\begin{equation}
\label{eq:PL2d}
    \sigma_N^2(R) \simeq \frac{1}{4 \sqrt{\pi}} \left( \frac{R}{\delta} \right)^{2-\alpha} \left( \frac{2^{4-\alpha } \Gamma \left(\frac{1}{2}-\frac{\alpha }{2}\right)}{(\alpha +2) \Gamma \left(2-\frac{\alpha }{2}\right)} + \left(\frac{\delta }{R}\right)^{\alpha } \frac{\Gamma \left(\frac{1}{2}-\alpha \right) \Gamma \left(-\frac{\alpha }{2}-1\right) \Gamma \left(\frac{\alpha +1}{2}\right) }{\Gamma
   \left(\frac{3}{2}-\frac{\alpha }{2}\right) \Gamma \left(\frac{\alpha }{2}+2\right) \Gamma (-\alpha )} \right)
\end{equation}
while in $d=3$ we have, setting now $\rho = 3/(4\pi \delta^3)$
\begin{equation}
\label{eq:PL3d}
    \sigma_N^2(R) \simeq 3 \left( \frac{R}{\delta} \right)^{3-\alpha} \left(\frac{3\ 2^{1-\alpha }}{(\alpha -1) \left(\alpha ^2-9\right)} - \left(\frac{\delta }{R}\right)^{\alpha } \frac{9\ 2^{1-2 \alpha } (2 \alpha +1) \sin (\pi  \alpha ) \cos ^2\left(\frac{\pi  \alpha }{2}\right) \Gamma (1-\alpha )^2 \Gamma (2 \alpha -1)}{\pi  (2 \alpha -3) \left(\alpha ^2+4 \alpha +3\right)^2}\right)
\end{equation}
\end{widetext}

\begin{figure}
    \centering
    \includegraphics[width=3in]{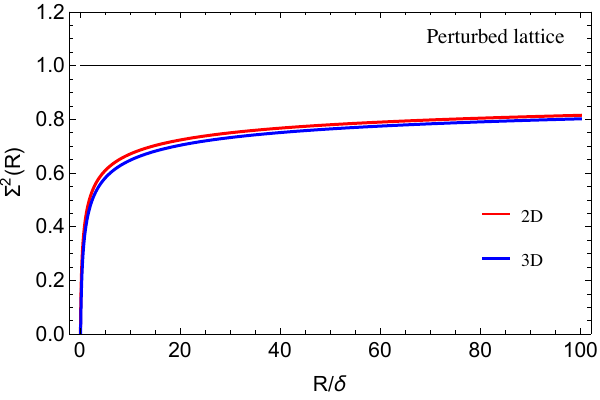}
    \caption{Scaled number variance for the 2D (blue) and 3D (orange) perturbed lattice, where we set $\alpha=1/4$ and $\delta=1$, according to Eqs.~\eqref{eq:PL2d},~\eqref{eq:PL3d}. In both cases, we also scaled the number variance so that it asymptotes to $1$ for better comparison. The approach to the asymptotic value is slow, governed by a power-law $R^{-\alpha}$.}
      \label{fig:numvar_PL}
 \end{figure}

We find that both in 2D and 3D, the behavior of the scaled number variance is given by
\begin{equation}
\label{eq:Sigma_pert_lattice}
    \Sigma^2(R) \simeq 1 - c_d(\alpha) \left(\frac{\delta}{R}\right)^{\alpha}, 
\end{equation}
where $c_d$ is a $d$- and $\alpha$-dependent constant that can be easily obtained from Eqs.~\eqref{eq:PL2d},~\eqref{eq:PL3d} using the definition Eq.~\eqref{eq:Sigma_def}. The result for the scaled number variance Eq.~\eqref{eq:Sigma_pert_lattice} is shown in Fig.~\ref{fig:numvar_PL}.
 Therefore, not only the leading behavior of the number variance $\sigma_N^2(R)$, but also the subleading correction to the asymptotic scaling is governed by the exponent $\alpha$. The hyperuniform behavior at the largest scale is thus also connected to the local uniformity in perturbed lattice configurations.

\section{Conclusion and Discussion}
\label{sec:conclusions}

\begin{table*}
\centering
\begin{tabular}{@{}lccc@{}}
\toprule
\textbf{System} & \textbf{Class} & $\mathbf{d}$ & $\mathbf{1 - \Sigma^2(R)}$ \\
\midrule
Step $g_2$ & I & $2$ & $-1/(40R^2)$  \\
Step $g_2$ & I & $3$ &  $-1/(18R^2)$  \\
SHU $\chi \ll 1$ & I & $2$ & $ - \cos(2KR)/(2KR)$ \\
SHU $\chi \ll 1$ & I & $3$ & $ - \sin(2KR)/(2KR)$ \\
One-comp. plasma & I & $2$ &  $ - 1/(16\pi R^2)$  \\
One-comp. plasma & I & $3$ &  $ - 1/[16(\pi/6)^{2/3} \Gamma(1/3) R^2]$  \\
Gaussian $g_2$ & I & $1$ &  $ - \exp(-4\pi R^2)/(8\pi R^2)$  \\
Gaussian $g_2$ & I & $2$ &  $ - 1/(16\pi R^2)$  \\
Gaussian $g_2$ & I & $3$ &  $ - 1/(6\pi R^2)$  \\
Fermi sphere & II & $1$ & $ 4.108/\ln{R}$ \\
Fermi sphere & II & $2$ & $ 3.308/\ln{R}$ \\
Fermi sphere & II & $3$ & $ 2.173/\ln{R}$\\
Perturbed lattice $(\alpha = 0.25)$ & III & $2$ & $-0.587/R^{1/4}$\\
Perturbed lattice $(\alpha = 0.25)$ & III & $3$ & $-0.629/R^{1/4}$\\
\bottomrule
\end{tabular}
\caption{Comparison of the corrections to the scaled number variance $1 - \Sigma^2(R)$ across various disordered hyperuniform systems discussed in the text.}
\label{table:summarize}
\end{table*}

In this work, we investigated the degree of uniformity of hyperuniform systems at small and intermediate scales by repurposing the local number variance $\sigma_N^2(R)$ that was used as an order metric in Ref.~\cite{Maher2024Local}. Specifically, we employ the scaled number variance $\Sigma^2(R)$, defined by Eq.~\eqref{eq:Sigma_def}, to facilitate the comparison of local uniformity across systems.
We examined class I, II, and III hyperuniform systems and computed -- in most cases analytically -- both the exact and asymptotic forms of the scaled number variance $\Sigma^2(R)$.

Our findings, which are summarized in Table~\ref{table:summarize}, are as follows. In class I disordered hyperuniform systems, presented in Sec.~\ref{sec:classI}, the subleading correction to the scaled number variance is an integer power of $1/R$. In particular, apart from SHU systems at small $\chi$ values, the contribution is $1/R^2$, therefore vanishing quickly and denoting a high degree of local uniformity. In the case of crystals and quasicrystals, the scaled number variance approaches the average value at distances comparable with the nearest neighbor spacing, and undamped oscillations around the average value $\Sigma^2(R) = 1$ persist at all length scales. 
In Sec.~\ref{sec:classII} we have discussed class II hyperuniformity, for which $\sigma_N^2(R) \sim R^{d-1}\ln R$. In this case, the subleading corrections are proportional to $1/\ln R$, thus signaling a lower degree of uniformity than class I. 
We then discussed class III hyperuniform systems in Sec.~\ref{sec:classIII}, showing that the exponent $\alpha$ governing the structure factor at small wave number is also responsible for the subleading corrections to the number variance, proportional to $R^{-\alpha}$. 
We note that reducing $\alpha$, thus degrading hyperuniformity at the largest scale, also reduces the effective uniformity across length scales, consistent with the expected behavior at $\alpha \to 0$, when the system ceases to be hyperuniform.

Note that our findings quantify the qualitative features that can be observed in Fig.~\eqref{fig:patterns}. In particular, class I hyperuniform systems show a higher degree of local uniformity, as quantified by $\Sigma^2(R)$.
Moreover, class III configurations (see Fig.~\ref{fig:patterns}(i)) are effectively more uniform at intermediate scales compared to class II (see Fig.~\ref{fig:patterns}(h)), despite having a weaker form of hyperuniformity. This observation is captured by the behavior of the scaled local number variance (cf. Table~\ref{table:summarize}).
There are, however, cases for which some more care is needed. Consider, for example, Fig.~\ref{fig:patterns}(e) and~\ref{fig:patterns}(i), representing, respectively, a class I and class III system. Qualitatively, the class III system in Fig.~\ref{fig:patterns}(i) seems more uniform than the SHU system (class I) in Fig.~\ref{fig:patterns}(e). This apparent issue is due to the value of $\sigma_N^2(R)$ in the two systems, and at what length scale they are comparable in the two systems. In particular, the class III configuration has been generated at unit number density, and, using Eqs.~\eqref{eq:PL2d} and~\eqref{eq:SHU_2d}, a direct comparison shows that the class III model has a lower number variance than the class I at sizes comparable to the ones displayed in Fig.~\eqref{fig:patterns} and smaller. This can also be understood from Fig.~\ref{fig:num_var_SHU}, showing that $\sigma_N^2(R)/R$ saturates at a large value compared to other systems. Therefore, it is important to specify that the information encoded in $\Sigma^2(R)$ can discern different degrees of uniformity between systems having a similar leading asymptotic behavior.

The results presented in this work are based on explicit calculations performed on specific models. Since our investigation covers nearly all known models of hyperuniform systems, we expect the conclusions to hold at least for this broad class of systems. 
It is natural to ask whether it is possible to find or design hyperuniform systems with anomalous subleading contributions to the scaled number variance, {\it e.g.} a class I system such that $1-\Sigma(R) \propto 1/\sqrt{R}$ or $1-\Sigma(R) \propto 1/\ln{R}$. We leave this question for future work.

The approach and the results presented in this work open the way to interesting future research directions. One of them is the possibility of defining a “town theorem" for hyperuniform systems. Namely, for a region of area (or volume) $\Omega$, how far away is a region with the same configuration up to rotations and reflections and within a specified bounded error? It is clear that, for crystals and quasicrystals, such a distance is very small, comparable with the nearest neighbor distance. It would be interesting to address how the distance $r$ changes in disordered hyperuniform systems: for instance, one would naively expect that for SHU systems at large $\chi$, $r$ is not very large (see Fig.~\ref{fig:patterns}(f)), and that it should be possible to relate this problem to the local uniformity discussion presented in this work.

Our findings are also relevant for more practical purposes. Experiments and numerical simulations are limited in system size, and our findings provide a way to determine which finite sizes are sufficient to probe the desired hyperuniform property of the system under consideration. For instance, in class I systems, it is often sufficient to have rather small systems, while considerably larger samples are required in the case of class II and III systems to probe their hyperuniform nature adequately.

In addition, the local uniformity of disordered hyperuniform systems is relevant for their physical properties. In particular, the suppression of volume-fraction fluctuations, corresponding to enhanced local uniformity, is known to reduce the propagation of cracks in disordered hyperuniform systems \cite{To00a}. 
A further suggestion comes from the fact that designed hyperuniform composites have been shown
\cite{xu_microstructure_2017} 
to possess a significantly higher brittle fracture strength than nonhyperuniform ones. We also expect that local uniformity properties can lead to advantages in transport and optical properties. This expectation is supported by the fact that, in stealthy hyperuniform systems, it has been shown in a series of recent works
\cite{kim_effective_2023,kim_accurate_2024,Vanoni2025Dynamical,To21d} 
that diffusion spreadability and optical transparency are enhanced compared to other hyperuniform and nonhyperuniform systems. 
We expect that a high degree of local uniformity can be instrumental for improved transport and optical properties, also in non-stealthy systems. 
We leave the investigation of such questions for future work.

\acknowledgments

The authors are grateful to Jaeuk Kim, Murray Skolnick, and Haina Wang for their help in generating some of the panels in Fig.~\ref{fig:patterns}.
This work was supported by the Army Research Office
under Cooperative Agreement No. W911NF-22-2-0103.

\section*{Data availability}

The data underlying the results presented in this paper are not
publicly available at this time, but may be obtained from the
authors upon request.

\bibliography{references.bib}

\end{document}